# Ionic liquid–electrode interface: classification of ions, saturation of layers, and structure-determined potentials

Karl Karu [1], Eva Roos Nerut [1], Xueran Tao [1], Sergei A. Kislenko [2], Kaija Pohako-Esko [3], Iuliia V. Voroshylova [4]*, Vladislav B. Ivaništšev [1]*

[1] Institute of Chemistry, University of Tartu, Ravila 14a, 50411 Tartu, Estonia

[2] Joint Institute for High Temperatures, Russian Academy of Sciences, Moscow, 125412 Russia

[3] IMS Lab, Institute of Technology, University of Tartu, Nooruse 1, 50411 Tartu, Estonia

[4] REQUIMTE LAQV, Department of Chemistry and Biochemistry, Faculty of Sciences, University of Porto, 4169-007 Porto, Portugal

* Correspondence: vladislav.ivanistsev@ut.ee, voroshylova.iuliia@fc.up.pt

**Abstract**

Progress in electrochemical applications of ionic liquids builds on an understanding of electrical double-layer. This computational study focuses on structure-determined quantities – maximum packing density, potentials and capacitances – evaluated using a one-electrode electrical double-layer model. Interfaces of 40 studied ions are grouped into four distinct classes according to their characteristic packing at the model surface. The simulations suggest that the exact screening by a monolayer of counter-ions (preceding the crowding of ions) is unlikely for common ions within the electrochemical stability window of corresponding ionic liquids. This work discusses how the assessed structure-determined quantities can guide the experimental tuning of (electro/mechano)chemical properties and characterising the structure of ionic liquid–electrode interfaces.

## 1.1. Introduction

The potential of applying ionic liquids (ILs) in technological applications motivates extensive studies of the electrical double layer (EDL) at IL–electrode interfaces [1]. Remarkable advances have been reached in the experimental characterisation of such interfaces. Namely, the unique layering of ions near the electrode surface (known as *overscreening*) was accessed with an impressively wide range of spectroscopy [2–10], microscopy [11–17], and electrochemistry [18–26] techniques. In addition to overscreening, the theoretically predicted *crowding* of counter-ions [27–30] was observed in computer simulations [31–33]. The modelled overscreening-to-crowding transition corresponds to the maximum density of ions packed in a monolayer of counter-ions [34–36], which appears in various molecular dynamics (MD), Monte Carlo, and classical density functional theory studies [37–39]. However, neither crowding nor monolayer formation has been decisively demonstrated in experiments referring to the overscreening and crowding concepts [40–48].

In a series of electrochemical experiments, Belotti *et al.* recently applied a voltage pulse and attributed the induced long-living EDL structure to a crowding regime [40]. Using atomic force microscopy, Jurado *et al.* observed a change in the EDL thickness and attributed it to the overscreening-to-crowding transition [41]. Dutta *et al.*, similarly to previously mentioned researchers, (mis)used the term "crowded" to describe the variation in X-ray reflectivity – a method that does not directly differentiate between anions and cations – actually showing that the described dense EDL is a mixture of both counter- and co-ions [42–44]. Klein *et al.* also associated the decrease in differential capacitance (measured by impedance spectroscopy) with crowding [45]. In all these studies, no structural evidence supporting crowding was provided. Nishi *et al.* applied various techniques to access the structure and quantify the absence of co-ions in the contact layer – a situation when only counter-ions are in contact with the surface – which they labelled crowding [46–48]. Similarly, in Joint density functional theory, Ma *et al.* reused the term "crowding" for the reorientation of ions in a saturated contact layer, preceding the true crowding [49]. Below, we show that such saturation corresponds to the overscreening regime. Thus, in these and similar studies, the term "crowding" is probably misused to describe fascinating yet distinct phenomena. The question is: How to discuss new and anomalous phenomena in terms of overscreening and crowding, even when comprehensive and complementary experimental evidence is absent? The present work suggests the maximum packing density of IL ions ($\theta_M$) and the Galvani potential of monolayer formation ($\varphi_M$) as a ballpark estimate for electrochemically measurable surface charge density ($\sigma$) and applied potentials ($U$ vs. potential of zero charge (PZC)) for avoiding ungrounded speculations.

Previously, $\theta_M$ was evaluated for a limited set of ions in work [49,50], concluding that the monolayer of practically important ions (like $Im_4^+$, $TFSI^-$, $FEP^-$; see the SI Table S1) is unlikely to form under experimental conditions. In this work, we predicted $\theta_M$ values for a larger set of ions with different sizes and shapes (see Figure 1). This article describes how similar structure-determined quantities can guide the experimental characterisation of crowded EDL and tuning (electro/mechano)chemical properties of IL–electrode interfaces.

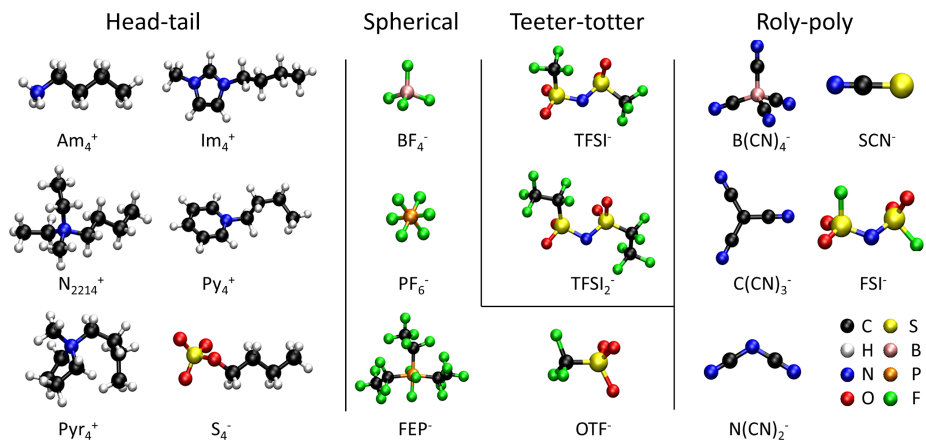

Figure 1: Ball-and-stick model of selected ions representing four identified classes. See Appendix Table 1. for names and acronyms.

### 1.2. Methods

#### 1.2.1. Phenomenological model

The phenomenological ionic bilayer model simplifies the layered IL–electrode interface to two layers so that the contact layer of counter-ions overscreens the surface charge and the subsequent layer neutralises the charge excess [36]. This model defines structure-determined potentials ($\varphi_S$ and $\varphi_M$) and capacitances ($C_S$ and $C_M$) in terms of a scaling exponent ($a$), the Helmholtz capacitance ($C_H$), relative and vacuum permittivities ($\varepsilon$ and $\varepsilon_0$), the distance between the EDL charge density planes ($l$) and the maximum packing density ($\theta_M$) as:

$$C_S = C_H \text{ and } C_M = aC_H \text{ and } C_H = \varepsilon\varepsilon_0/l \qquad (1)$$

and

$$\varphi_S = \frac{\theta_M}{C_H}\left(\frac{1}{a}\right)^{\frac{1}{a-1}} \text{ and } \varphi_M = \frac{\theta_M}{C_H} \qquad (2)$$

The derivation of Eqs. 1 and 2, along with a description of all quantities, is included in the SI. The evaluated $\theta_M$ and estimated ($C_M$, $\varphi_M$) and ($C_S$, $\varphi_S$) values for 40 ions are presented below in Tables 1–3. $\varphi_M$ is the potential at which the monolayer of counter-ions exactly screens the surface charge ($\sigma = -\theta_M$). $\varphi_S$ is the potential at which the contact layer becomes saturated with counter-ions. In between $\varphi_S$ and $\varphi_M$, the prevailing mechanism of surface charge screening is the desorption of the co-ions from the second layer. This mechanism leads to a decrease in differential capacitance. $C_S$ and $C_M$ are the differential capacitance values at $\varphi_S$ and $\varphi_M$. In this model, the distance of the closest approach gives the Helmholtz capacitance ($C_H$).

*1.2.2. Simulated model*

The EDL in ILs is commonly simulated using a two-electrode model with IL confined between electrodes [51,52]. The two-electrode model allows for keeping constant surface charge, potential difference, or applied field in the simulations [53–55]. The bisection of a two-electrode model gives a one-electrode EDL model that is commonly applied in electrocatalysis simulations [56]. Figure 2 illustrates charging regimes, structures, and mechanisms showing only a minimal number of counter- and co-ions that screen the surface charge in both models (see Refs. [28,35] for details).

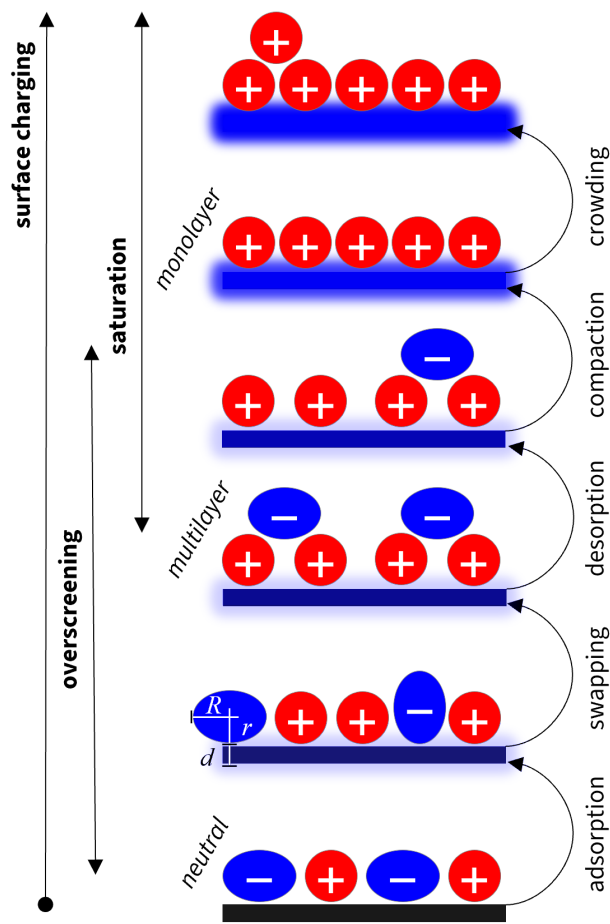

Figure 2: Schematic representation of an ionic liquid–electrode interface at variable surface charge density that is screened by shown counter- and co-ions. Processes of surface charing, overscreening, and saturation are highlighted with bold on the left. Charging mechanisms are marked on the right. Neutral, multilayer, and monolayer structures are labelled in italics. Parameters *d*, *r*, and *R* are shown.

This work simplifies the one-electrode EDL model to include only counter-ions, as in the Helmholtz model [57]. This model estimates the maximum packing density, assuming that $\theta_M$ is determined by the geometry of ions. Previous simulations show that when $\sigma = -\theta_M$, the monolayer of counter-ions completely screens the surface charge density [35]. On this basis, it can be assumed that the $\theta_M$ and $\varphi_M$ values are similar for simulations with both one-electrode and more sophisticated two-electrode models. This assumption is valid for the packing of $PF_6^-$ simulated in this work and Refs. [36,58] (see Figure 3) using one- and two-electrode models, respectively. Herewith, the simplified one-electrode model shortens the simulation time by two orders of magnitude compared to the two-electrode model.

### 1.2.3. Molecular dynamics simulations

The simulations were performed with the GROMACS 2020 software [58–64] and OPLS-AA force fields with ±1$e$ charges on ions [65–67]. The graphene-like electrode was prepared using the Atomic Simulation Environment [68] with a size of 4.26 nm × 3.94 nm (640 carbon atoms) and a 0.30 nm tolerance for packing. The Packmol software [69] generated five replicas with different packing for a variable number of counter-ions.

First, a rough energy minimisation run was performed with a steepest descent method with 0.01 nm step size and 10 kJ mol$^{-1}$ nm$^{-1}$ force convergence threshold. Then, an equilibration step was run for 20 ps with 0.5 fs timestep, generating velocities corresponding to the temperature of 100 K. The production run was done in the Canonical ensemble (*NVT*) for 1000 ps with 1 fs time step and velocities corresponding to 353 K. The Verlet cut-off scheme was employed, and *xy*-periodic boundary conditions were applied with the particle-mesh Ewald and 3dc Ewald correction [70–72]. Next, the trajectories were analysed. Charge and mass density profiles, as well as electrostatic potential profiles, were obtained using GROMACS tools and processed using Python scripts. MDanalysis software was used to analyse the distances [73,74]. VMD software [75] was employed for visualisation.

### 1.2.4. Density functional theory calculations

Density functional theory calculations were performed to estimate the reduction and oxidation potentials of the IL ions. The calculations were performed using the r$^2$SCAN-3c method [76] as implemented in the Orca 5.0.2 quantum chemistry program package [77]. The r$^2$SCAN-3c is a composite method based on a meta-generalized-gradient approximation density functional [78,79], which employs a custom triple-zeta Gaussian basis set [76], D4 dispersion correction [80,81], and geometrical counterpoise correction [82]. The SCAN functional demonstrated great accuracy for determining the energetic parameters of IL ions, especially when paired with dispersion and counterpoise corrections. [83]

The geometries of the ions were optimised using nominal charges of +1e for cations and −1e for anions. Furthermore, single-point calculations were performed to calculate the ionisation potential (IP) and electron affinity (EA) using these geometries with ionic charges of 0e and +2e for cations and −2e and 0e for anions. Under the assumption that oxidation and reduction occur adiabatically and are not affected by the environment, the IL electrochemical window for each anion–cation combination was estimated as

$$EW = \min(IP) + \max(EA), \qquad (3)$$

where the EA is defined as the energy released upon accepting one electron.

*1.2.5. Workflows*

The NaRIBaS (Nanomaterials and Room-temperature ILs in Bulk and Slab) scripting framework was employed to construct flexible and automatic workflows [64]. The scripts with the topology files and input parameters are available at Git-Hub [84,85].

*1.2.6. Structure-determined packing density, potential drops, and capacitances*

The contact layer density ($\theta$) was extracted from number density profiles using the numerical evaluation of the density slices. Moving from the electrode to the opposing end of the simulation box, the beginning of the contact layer was defined as a point where the density in a slice exceeds an ion-specific threshold. The density of a contact layer was found by integrating the density of the slices in between. Figure 3 illustrates that the maximum packing density ($\theta_M$) value corresponds to the plateau at $|\theta|$ vs. $|\sigma|$ plot. The results were visually checked with VMD software for the crowding condition.

Figure 3 also shows the agreement between $\theta_M$ estimation procedures using the one- and two-electrode models [36,58]. Note that $\theta_M$ is reached at lower $\sigma$ values in the two-electrode simulations. First, this is due to the presence of co-ions after the contact layer of counter-ions, *i.e.* overscreening. Second, the inflexion point corresponds to the saturation potential ($\varphi_S$), after which the prevailing mechanism of the surface charge screening is the desorption of co-ions from the EDL. Most importantly, this study shows that in both procedures, the exact screening (by the monolayer of counter-ions) is reached when $\sigma = -\theta_M$.

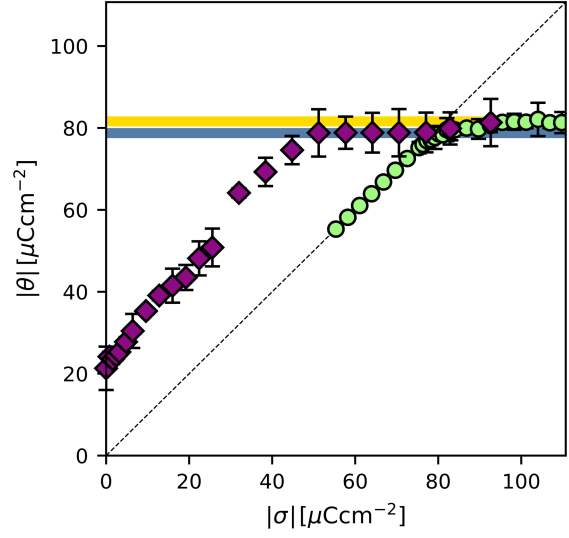

Figure 3: Contact layer charge density ($\theta$) dependence on the surface charge density ($\sigma$) for $PF_6^-$. The maximum packing density ($\theta_M$) corresponds to the plateau. The number of $PF_6^-$ ions in the contact layer is shown with circles and diamonds for simulations with one and two electrodes, respectively. Data for the two-electrode model is from Refs. [36,58].

The monolayer potential drop ($\varphi_{M,P}$) was calculated using the Poisson equation and the charge density profiles at estimated $\theta_M$. Then, these $\varphi_{M,P}$ values were recalculated under two assumptions: first, the partial charge transfer can be accounted for by varying relative permittivity ($\varepsilon_r$); second, the total potential drop can be split into electrode and electrolyte contributions.

The first assumption views the partial charge transfer as an extreme of electronic polarisation in the EDL. $\varepsilon_r$ values from 1.6 to 2 are commonly used in MD simulations to account for electronic polarisation through atomic charge scaling as $q = \varepsilon_r^{-1/2}$ [59]. In these simulations, $\varepsilon_r = 1$ was used and then rescaled during analysis.

The second assumption was tested by comparing $\varphi_M$ values evaluated directly *via* the Poisson equation ($\varphi_{M,P}$) and using simulated $\theta_M$ in the following expression with $q = \pm 1$ and $\varepsilon_r = 1$:

$$\varphi_{M,P} = -\frac{d+r}{\epsilon_r \epsilon_0}\theta_M = -\frac{d\theta_M}{\epsilon_r \epsilon_0} + \frac{r}{R} \cdot \frac{q}{\epsilon_r \epsilon_0}\sqrt{\frac{\theta_M}{4q}} \qquad (4)$$

where $d$ is the surface atom's radius, $q$ is the counter-ion charge (±1e in this study), $r/R$ is the compression factor of an ion, *i. e.* the ratio between contact ($r$) and lateral ($R$) radii of the counter-ion (see Figure 2). The lateral radius was estimated as $R^2 = ¼·q/\theta_M$. The compression factor indicates the degree of deviation of the ion's elliptic shape from a spherical form. The sum of $d$ and $r$ assumes that the distance between the surface and the contact layer is determined by the radii of surface atoms and counter-ions. This assumption is supported by the high coefficients of determination ($R^2$) of 0.987 for linear regression of $\varphi_{M,P}$ values (see SI Figure S1). Indeed, the potential drop within distance $d$ from the electrode can be associated with the electrode and corrected as described in Ref. [58]. So, $d$ is set to 0 nm in Eq. (4) under the assumption that the surface charge plane of an ideal metal electrode extends by one-half of an interplanar spacing as follows from theory [60,61] and calculations [62,63]. The $\varphi_M$–$C_M$ and $\varphi_S$–$C_S$ pair values given below are calculated from $\theta_M$ and $r$ values using Eqs. 1–2 with $l = r$, $d = 0$, and $\varepsilon_r = 1.6$. These parameters provide a lower estimate for the $\varphi_M$–$C_M$ and $\varphi_S$–$C_S$ pair values.

**1.3. Results**

*1.3.1. Taxonomy of counter-ions*

The examined ions were grouped into four classes according to their packing features: ball, head-tail, roly-poly, and teeter-totter. The $-\theta(\sigma)$ dependence differentiated the interfacial behaviour, as in Figure 3. The ball and teeter-totter ions give a simple $|\theta|$–$\sigma$ plot with two clear intersecting lines, while head-tail and roly-poly ions give more complex $|\theta|$–$\sigma$ plots due to the reorientation of individual ions.

*1.3.2. Ball class*

As the name suggests, the ball class includes spherical ions such as $BF_4^-$, $Br^-$, $Cl^-$, $FEP^-$, $I^-$, $OTF^-$ and $PF_6^-$. Their main packing feature is that they behave like coarse-grained spheres during the simulations and form close-packed structures.

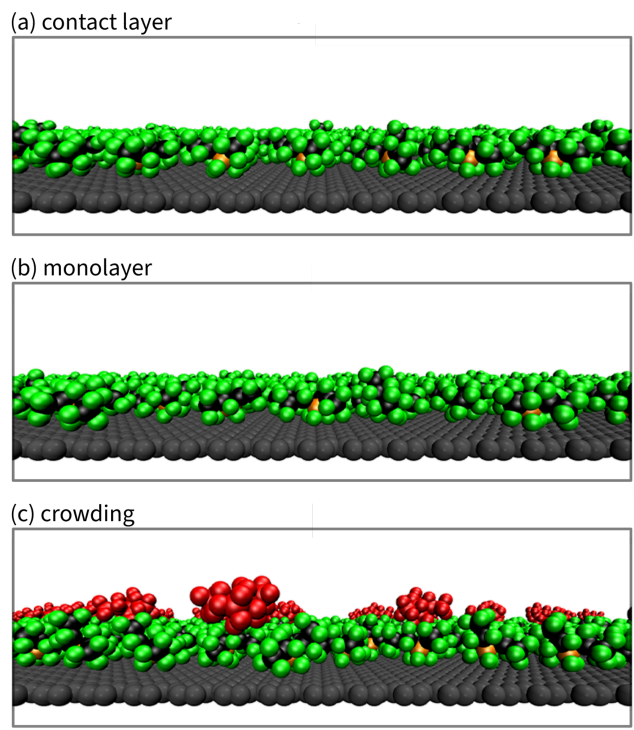

Figure 4: Side view on the ball-and-stick models of the FEP⁻ | electrode interface. The contact layer (a) and the monolayer (b) are visually indistinguishable, yet the crowding state (c) stands out due to ions in the second layer (shown in red).

Figure 4 shows the packing of ball FEP⁻ ions. Due to the simple shape, ions are well organised and fill the surface in a highly ordered manner, resulting in dense packing. Ions in this class do not change orientations upon increasing $|\sigma|$. Thus, the contact layer looks indistinguishable from the monolayer (compare Figure 4a and Figure 4b) and is distinct from the crowded double layer (see Figure 4c). In the crowding regime, the projection of two ionic layers resembles a Moiré pattern (see Figure 5a), like in simulations of coarse-grained ionic liquids in Ref. [34]. Moreover, a characterization structural transition is a stepwise increase of $\theta$ with increasing $|\sigma|$. In Figure 5, $\theta$ increases sharply from $-150$ to $-162$ µC·cm⁻², which is disproportionate to the $\sigma$ change of only 1 µC·cm⁻². Analogeous phase transition is observed experimentally for halide anions that adsorb specifically from aqueous solutions [70]. Moreover, analogues stepwise densening was also seen in MD simulations of Au(*hkl*) | BMImPF$_6$ interfaces [86].

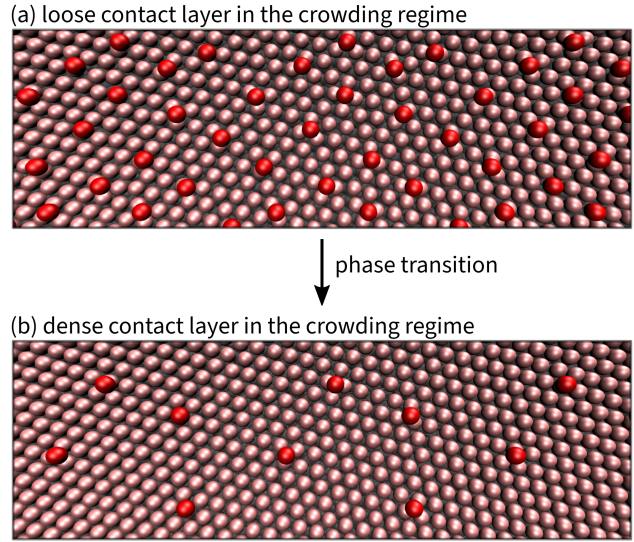

(a) loose contact layer in the crowding regime

phase transition

(b) dense contact layer in the crowding regime

Figure 5: Top view on Br⁻ adlayers in the crowding regime. Ions in the first layer are shown in pink (lighter), and in the second layer, they are shown in red (darker). The phase transition from a looser contact layer (a) to a denser contact layer (b) occurs upon adding a single Br⁻ anion to the model unit cell.

Table 1 summarises the simulated and estimated parameters of the modelled IL–electrode interfaces with ball ions. This data is discussed below. Note that in the case of $PF_6^-$ and $FEP^-$ the scaling exponent ($a$) is close to its geometrical value of 0.38 for packing of balls [36].

Table 1: Structure-determined parameters of the modelled EDLs with ball ions. The standard deviation, indicated in brackets, is computed based on the variation observed across multiple replicates.

| Ion | $\theta_M$ [µC·cm⁻²] | $\varphi_M$ [V] | $\varphi_S$ [V] | $C_M$ [µF·cm⁻²] | $C_S$ [µF·cm⁻²] | $a$ |
|---|---|---|---|---|---|---|
| $BF_4^-$ | −105.2(8) | 8.3 | 1.5 | 3.9 | 12.6 | 0.31 |
| $Br^-$ | −149.9(7) | 7.6 | 1.2 | 5.1 | 19.7 | 0.26 |
| $Cl^-$ | −181.3(25) | 7.7 | 1.2 | 5.5 | 23.6 | 0.23 |
| $FEP^-$ | −25.8 | 7.7 | 1.2 | 5.5 | 23.6 | 0.37 |
| $I^-$ | −87.8(19) | 6.0 | 1.0 | 3.8 | 14.8 | 0.26 |
| $OTF^-$ | −73.3(15) | 6.2 | 0.9 | 2.8 | 11.8 | 0.23 |
| $PF_6^-$ | −77.5(4) | 8.5 | 1.7 | 3.2 | 9.1 | 0.35 |

*1.3.3. Head-tail class*

Head-tail ions consist of an alkyl chain tail and a head, holding the ionic charge. For $Py_n^+$ and $Im_n^+$ the head is an aromatic cycle. For $Am_n^+$, $Pyr_n^-$ and $N_{2214}^+$ ions, the head is an (alkyl)ammonium group.

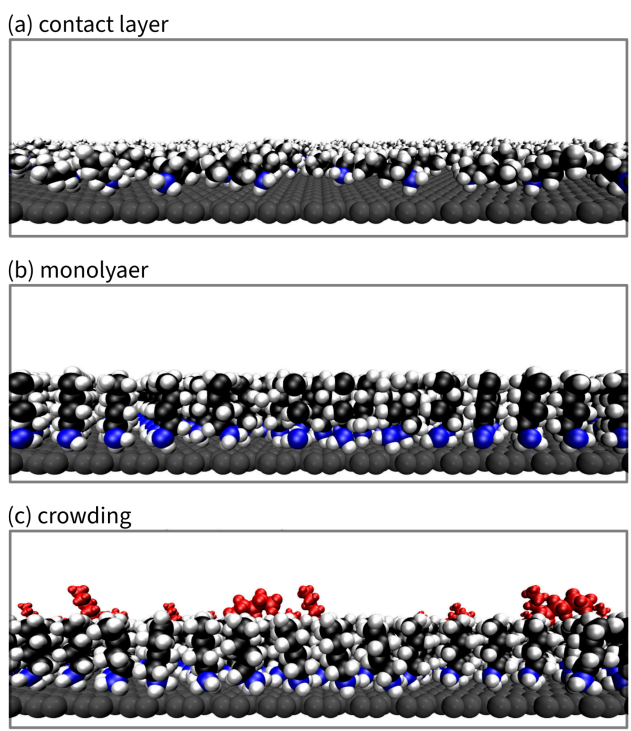

Figure 6: Side view on the ball-and-stick models of the $Am_4^+$ | electrode interface. The maximum packing density is reached in simulations of 103 $Am_4^+$ ions per electrode area. Ions of the second layer are shown in red.

Figure 6 shows layers of head-tail $Am_4^+$ ions. While at low $|\sigma|$, with enough voids on the surface, $Am_4^+$ tails lie parallel to the surface (Figure 6a), at high $|\sigma|$, $Am_4^+$ tails are forced to "stand up", *i. e.* point outwards the surface to free void and facilitate surface charge screening by $Am_4^+$ heads (Figure 6b). A similar phenomenon was described in detail from MD simulations of EDLs with $N_{1114}^+$, $N_{1144}^+$, $N_{1444}^+$, and $N_{4444}^+$ ions [48]. In the monolayer, all $Am_4^+$ tails are oriented outwards from the surface. Similar pictures are seen for all studied head-tail ions: first, in the overscreening regime, their tails can reorient; second, before and in the crowding regime, the EDL structure depends on the head-tail form and size. Five scenarios can be outlined:

(1) **Head reorientation:** $Im_n^+$ and $Py_n^+$ ions' aromatic rings (heads) change orientation from parallel to perpendicular relative to the surface. For example, the $Im_n^+$ monolayer has three distinguishable orientations of the $Im^+$ ring: one parallel and two perpendicular to the surface (with the $C^2$ atom pointing towards and outwards the surface). Such orientation resembles lying, standing downwards, and standing upwards at the surface and was observed in numerous MD simulations and experimental works [72,87–90].

(2) **Head-head stacking:** Ions that do not fit their head into the contact layer occupy voids between tails beyond the contact layer heads, thus appearing as head-to-head stacking.

(3) **Head-tail intercalation:** Ions that do not fit their head into the contact layer occupy voids between tails at a distance from the contact layer heads, thus appearing to intercalate the layer of tails.

(4) **Heads over tails crowding:** Ions that do not fit their head into the contact layer situated right beyond the layer of tails. In [Figure 6c](), such ions can not penetrate the layer of tails due to a very dense packing of $Am_n^+$ in the contact layer. The longer the tail, the denser the layer of tails. Thus, even if head-head stacking and head-tail intercalation are possible for ions with short tails and large heads, a larger fraction of ions crowd in the layer following the tails when the tails become longer.

(5) **Tail-tails stacking:** Such behaviour, described in "Monolayer to Bilayer Structural Transition in Confined Pyrrolidinium-Based Ionic Liquids" by Smith *et al.* [91–93], was not observed in the presented simulations. Unlike the experiments, where such transition was induced by increasing the number of carbon atoms in alkyl chains (*n*) from 8 to 10, the presented simulations might not reproduce that bilayer structure due to the absence of co-ions stabilising the transition.

Table 2 summarises simulated and estimated parameters of the modelled EDLs with head-tail ions with a tail length of 4 carbon atoms. Additional data is given in the SI Table S1. Ma *et al.* also simulated $Im_4PF_6$ with a two-electrode model and suggested $\theta_M(Im_4^+)$ of 40 $\mu C \cdot cm^{-2}$ [49], which agrees well with this and [50,90] studies.

Table 2: Structure-determined parameters of the modelled EDLs with head-tail ions. Only one set is shown per each family of ions, while more detailed data can be found in the SI Table S2. The standard deviation, indicated in brackets, is computed based on the variation observed across multiple replicates.

| Ion | $\theta_M$ [$\mu C \cdot cm^{-2}$] | $\varphi_M$ [V] | $\varphi_S$ [V] | $C_M$ [$\mu F \cdot cm^{-2}$] | $C_S$ [$\mu F \cdot cm^{-2}$] | $a$ |
|---|---|---|---|---|---|---|
| $AM_4^+$ | 98.3(12) | −3.5 | −1.0 | 16.9 | 28.1 | 0.60 |
| $IM_4^+$ | 44.9(7) | −3.3 | −0.8 | 6.8 | 13.6 | 0.50 |
| $Py_4^+$ | 42.4(9) | −3.0 | −0.9 | 9.8 | 14.2 | 0.69 |
| $Pyr_4^+$ | 44.9(10) | −5.3 | −1.6 | 6.0 | 8.4 | 0.71 |
| $N_{2214}^+$ | 45.6(4) | −6.2 | −1.3 | 2.8 | 7.4 | 0.38 |
| $S_4^-$ | 66.4(5) | 5.8 | 1.2 | 4.4 | 11.4 | 0.39 |

### *1.3.4. Roly-poly class*

As the class name suggests, roly-poly ions can change individual orientation from parallel to perpendicular to the surface upon increasing $|\sigma|$. Roly-poly ions "prefer" to lie flat on the surface at low $|\sigma|$, but as $|\sigma|$ increases to a certain point, ions begin to stand up to make more void for adsorbing ions. Figure 7a shows a particular state where all ions lie parallel to the surface. Increasing $|\sigma|$ forces ions to stand up, increasing the fraction of ions oriented perpendicular to the surface. In the monolayer of roly-poly ions, parallel and perpendicular orientations coexist, see Figure 7b. Moreover, roly-poly ions can orient towards and outwards the surface, *i.e.* the perpendicularly oriented $SCN^-$ ions contact the surface with sulphur or nitrogen atoms (see Figure 7b and Figure 7c). In general, roly-poly ions are drastically different from head-tail and ball ions in terms of packing at variable $|\sigma|$ (compare Figure 4, Figure 6, and Figure 7).

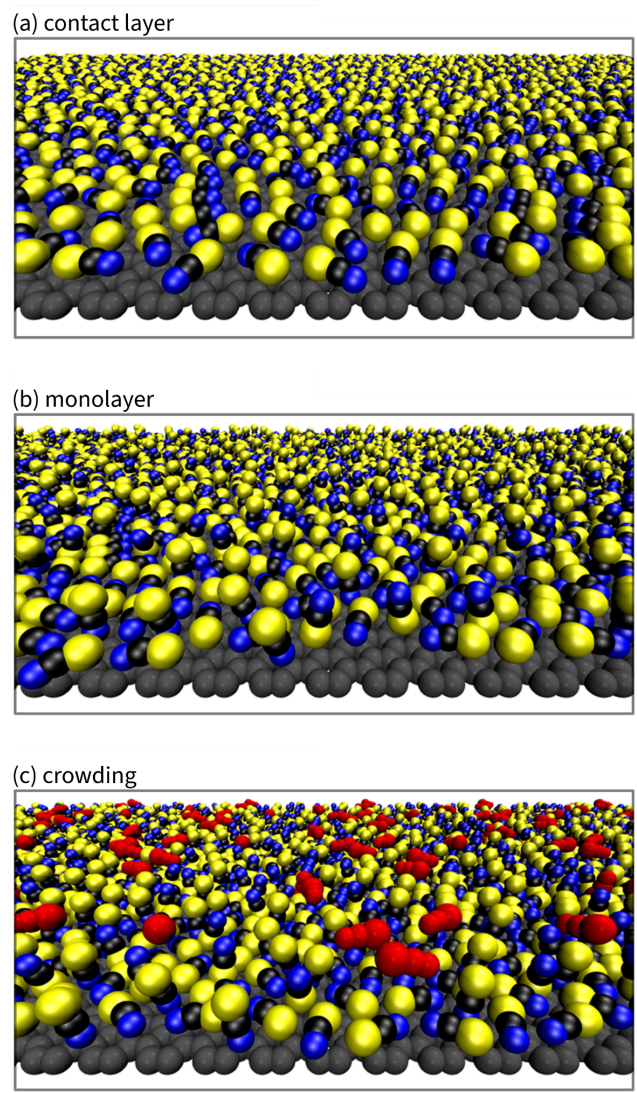

Figure 7: Side view on the ball-and-stick models of the SCN⁻ | electrode interface. In the low-density contact layer (a), SCN⁻ anions lie parallel to the surface. In the monolayer (b) and crowder state (c), a significant fraction of SCN⁻ ions orient perpendicular to the surface. The second layer is shown in red.

Table 3 summarises the simulated and estimated parameters of the modelled EDLs with roly-poly and teeter-totter ions. This data is discussed below.

Table 3: Structure-determined parameters of the modelled EDLs with roly-poly and teeter-totter ions. The standard deviation, indicated in brackets, is computed based on the variation observed across multiple replicates.

| Ion | $\theta_M$ [μC·cm$^{-2}$] | $\varphi_M$ [V] | $\varphi_S$ [V] | $C_M$ [μF·cm$^{-2}$] | $C_S$ [μF·cm$^{-2}$] | $a$ |
|---|---|---|---|---|---|---|
| B(CN)$_4^-$ | −64.7(15) | 9.1 | 2.4 | 3.8 | 7.1 | 0.53 |
| C(CN)$_3^-$ | −73.7(8) | 6.7 | 1.6 | 5.4 | 11.1 | 0.48 |
| N(CN)$_2^-$ | −108.4(15) | 7.3 | 1.2 | 3.8 | 14.8 | 0.26 |
| SCN$^-$ | −103.8(28) | 4.1 | 0.5 | 4.5 | 25.3 | 0.18 |
| FSI$^-$ | −59.0(8) | 5.7 | 1.1 | 3.4 | 10.4 | 0.33 |
| PFSI$^-$ | 30.9(5) | 2.8 | 0.6 | 4.9 | 11.1 | 0.44 |
| TFSI$^-$ | −42.2(4) | 3.7 | 0.9 | 5.7 | 11.4 | 0.50 |

### 1.3.5. Teeter-totter class

The teeter-totter ions class comprises ions with a particular structure, where one central atom connects to two similar groups on both sides.

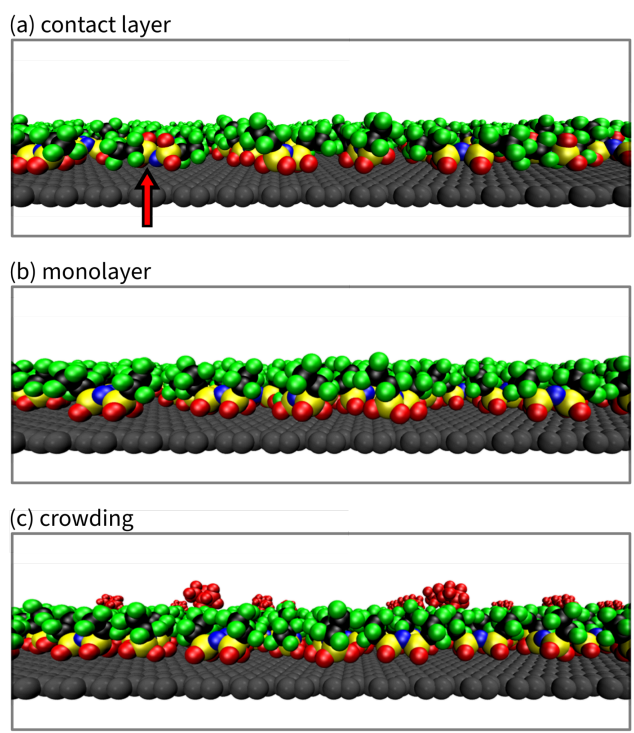

Figure 8: Side view on the ball-and-stick models of the PFSI$^-$ | electrode interface. The contact layer (a) contains trans-conformers of PFSI$^-$ (indicated with a red arrow), while in

the monolayer (b) all ions orient with oxygen atoms towards the surface, and in the crowding regime (c) additional ions (shown in red) form the second layer.

Although the teeter-totter ions can be considered linear and similar to the roly-poly class, they do not reorient with increasing $|\sigma|$. Only at very low $|\sigma|$, the teeter-totter ions might take a trans-configuration as marked in Figure 8a and in agreement with experimental results [47]. More negatively charged atoms of teeter-totter ions orient towards the surface at all studied $|\sigma|$.

The findings presented here align with previous MD simulations of ILs with $TFSI^-$ conducted by Sharma and Kashyap [94]. In close proximity to a positively charged graphene sheet, the major axis of symmetry of the $TFSI^-$ anions predominantly align parallel to the surface. This observation emphasises the significance of electrostatic interactions between oxygen atoms $TFSI^-$ in and the graphene sheet, as in this study.

### *1.3.6. Summary of results*

All evaluated $\varphi_{M,P}$–$\theta_M$ values along with the curve following Eq. (4) are shown in Figure 9. According to the compression factor ($r/R$), most of the ions are ellipses (like heads of $Am_n^+$ and $S_n^-$ with $r/R \approx 0.4$), some ions are slightly more spherical (halide anions and $Py_n^+$ with $r/R \approx 0.6$), while others are almost perfect spheres (cyano-based anions and $Pyr_n^+$ with $r/R \approx 0.8$). For ball ions, Figure 9 shows an obvious trend of rising $\varphi_M$ value with decreasing ion radius. In a clear example of halide ions, one can see that the smaller the ion, the higher the potential of the monolayer formation. This observation is *qualitatively* in line with predictions made by Ivaništšev and Fedorov [95], where the maximum packing density was approximated by $\theta_M = ¼·q/r^2$, taking $r = R$.

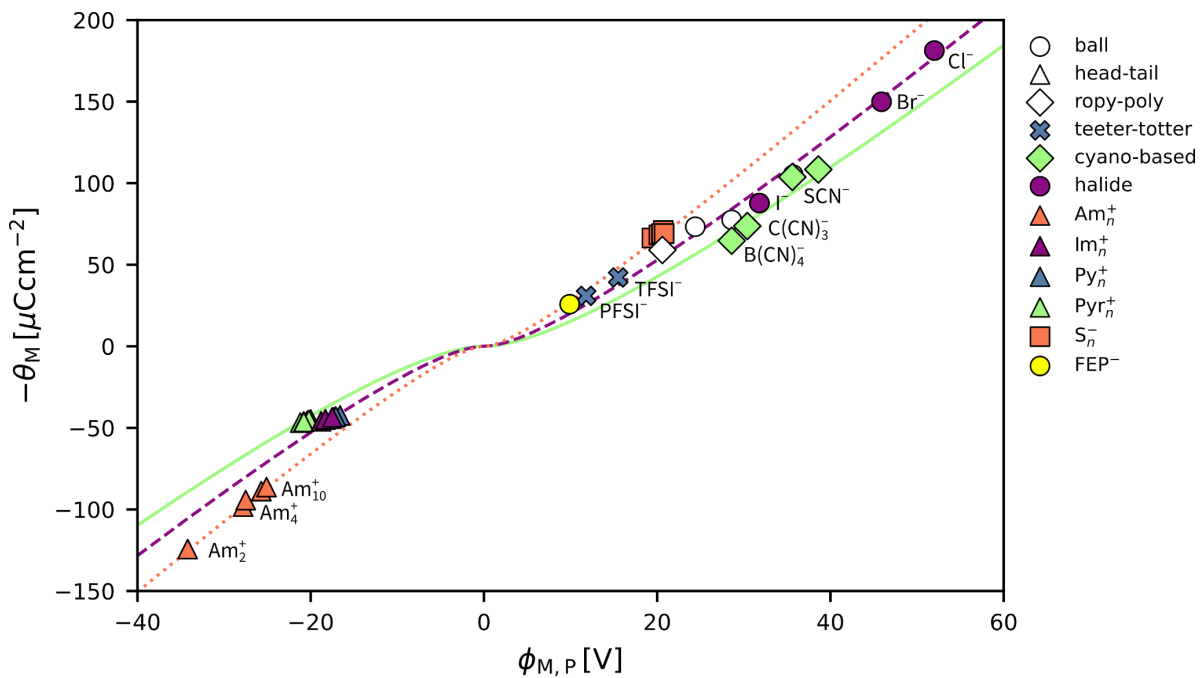

Figure 9: Calculated maximum packing density ($\theta_M$) and monolayer potential drop ($\varphi_{M,P}$) values. Lines follow Eq. (4) with compression factor ($r/R$) of 0.8 (solid), 0.6 (dashed), and 0.4 (dotted).

For head-tail ions, using radii from Refs. [73,74,96], Ivaništšev and Fedorov presumed a dependence of $\theta_M$ on the tail length [95]. However, in the present simulations, only $\theta_M$ of $Am_n^+$ depends on the tail length. Different from all simulated head-tail ions, in $Am_n^+$ the ammonium head and alkyl tail have similar cross-section diameters, so the stacking of tails determines the packing of heads, as can be seen in Figure 6. The lateral distance between tails increases with the number of bent conformations, *i.e.*, the tail lengthening. This effect is levelled down when the head is wider than the tail, like in $Im_n^+$, $Py_n^+$, and $Pyr_n^+$, yielding almost the same $\theta_M$ value for these ions (see Figure 9).

## 2. Discussions

### 2.1. Comparison of structure-determined potentials

Figure 10 summarises all corrected $\varphi_M$–$\theta_M$ values obtained with the Poisson equation, then (1) shifting the surface change plane by $-d\theta_M/\varepsilon_0$ with $d = 0.17$ nm and (2) dividing by $\varepsilon_r = 2$. The shown $\theta_M$ values correspond to the ion charge $q = \pm 1e$ as in Figure 9 and all Tables. Note that these values differ slightly from the $\varphi_M$ values calculated using Eq. (2) and shown in Tables 1–3 and SI Table S1.

The most marked feature is that all $\varphi_M$ values are outside the ±2 V vs PZC region, which is the experimentally measurable potential window [50]. In other words, true monolayer formation and crowding should not be experimentally reachable for all studied ions. This conclusion holds under the assumption that the partial charge transfer is below $1 - \sqrt{½} = 0.3e$. The highly polarisable $Py_n^+$ could form the monolayer at −2 V upon accepting $0.5e$ (see Figure 10). However, the scanning tunnelling microscopy study of $Py_4^+$ adsorption from $Py_4BF_4$ did not reveal any ordered structure [97]. Halide anions could form their monolayers at +2 V upon donating $0.7e$. Indeed, halide anions are well known for forming highly ordered adlayers even at lower relative potentials [78,98,99]. It is worth examining whether these adlayers meet the exact screening definition by checking for evidence of overscreening and crowding. For instance, by examining the composition of the second layer with XRD and STM [100].

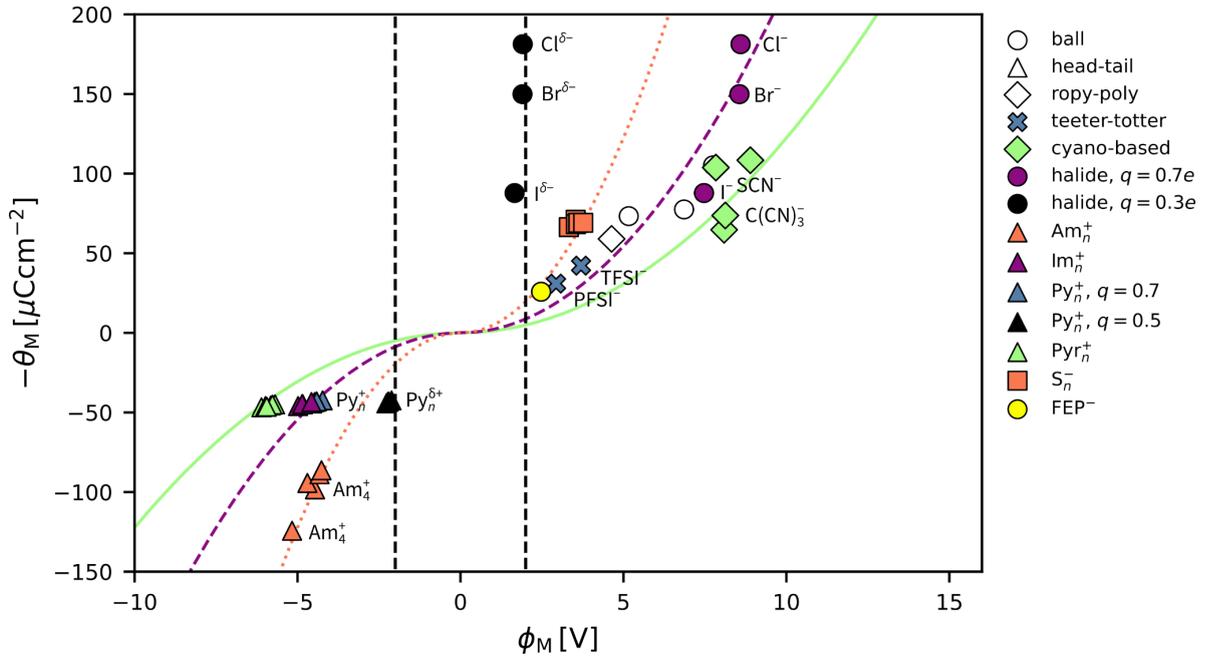

Figure 10: Simulated maximum packing density ($\theta_M$) and recalculated monolayer potential drop ($\varphi_M$) values. Lines follow Eq. (4) with $d = 0$, $\varepsilon_r = 2$, and compression factor ($r/R$) equals 0.8 (solid), 0.6 (dashed), and 0.4 (dotted).

On the contrary to $\varphi_M$ values, the estimated $\varphi_S$ values are within the ±2 V region, as reported in Tables 1–3. That means most of the reported ordered adlayers [11,16,24,79–81,83,101–103] form near the saturation potential ($\varphi_S$). Figure 2 illustrates this conclusion with results for two electrode simulations of $BMImPF_6$ [36,58], where $\varphi_S$ of 3 V was estimated for $PF_6^-$ anions. In this work, the corresponding estimation is 2.2 V for the same $\varepsilon_r$ of 1.6. Despite crude approximations in Eq. (2) for $\varphi_S$, it leads to a phenomenologically reasonable conclusion that interfacial processes attributed to crowding (e.g. Refs. [40]) are most probably occurring in the overscreening state (see Refs. [16,104]). Moreover, switching on and off applicable properties such as capacitance, lubricity, actuating property, [105–107] are probably related to the change in the screening mechanism at $\varphi_S$: from adsorption of counter-ions to desorption of co-ions from the second layer. [36,49,108]

## 2.2. Structure-determined potentials and the electrochemical window

A simple procedure for finding surfaces with $\varphi_M$ and $\varphi_S$ values within the electrochemical stability window (EW) of ionic liquid–electrode interfaces is a comparison of the lower estimate for $\varphi_M$ and $\varphi_S$ (Eqs. (1) and (2) with $l = r$, $d = 0$, and $\varepsilon_r = 2$) to the upper estimate for the EW (Eq. (3)). The latter assumes that ions decompose solely *via* electron transfer neglecting the solvation. The predicted EWs vary from 4 to 12 V for four hundred anion–cation combinations. These are overestimated due to ignorance of essential aspects such as EDL structure, decomposition mechanism, catalytic effect and reactivity of the surface, and partial charge transfer, which should narrow the EW. Moreover, adjustment by the PZC should further narrow the EW because anodic and cathodic processes are interrelated through the electroneutrality principle. Summing $\varphi_M$ and $\varphi_S$ for anion–cation combinations cancels $\varphi_{PZC}$ out. Tables 4 and 5 show the differences EW − $\Delta\varphi_M$ and EW − $\Delta\varphi_S$ for selected anion–cation ILs. The full compassion (given in the SI) provides an optimistic estimation of whether $\varphi_M$ and $\varphi_S$ fit into the EW.

Table 4: Difference between predicted EW and $\Delta\varphi_M$ values. Positive values imply that $\varphi_M$ values might be experimentally reached.

|  | Br$^-$ | Cl$^-$ | I$^-$ | BF$_4^-$ | PF$_6^-$ | N(CN)$_2^-$ | C(CN)$_3^-$ | B(CN)$_4^-$ | FSI$^-$ | TFSI$^-$ | PFSI$^-$ | FEP$^-$ |
|---|---|---|---|---|---|---|---|---|---|---|---|---|
| Pyr$_4^+$ | −8.4 | −8.5 | −6.9 | −5.8 | −5.7 | −7.7 | −7.0 | −6.0 | −5.4 | −3.4 | −2.5 | −0.9 |
| AM$_4^+$ | −5.3 | −5.4 | −3.7 | −2.7 | −2.5 | −4.5 | −3.9 | −2.8 | −2.3 | −0.3 | 0.6 | 2.2 |
| IM$_4^+$ | −4.4 | −4.5 | −2.8 | −1.8 | −1.6 | −3.6 | −3.0 | −1.9 | −1.4 | 0.6 | 1.5 | 3.2 |
| Py$_4^+$ | −2.6 | −2.7 | −1.0 | 0.0 | 0.2 | −1.8 | −1.1 | −0.1 | 0.4 | 2.4 | 3.3 | 5.0 |

Table 5: Difference between predicted EW and $\Delta\varphi_S$ values. Positive values imply that $\varphi_S$ values might be experimentally reached.

|  | Br$^-$ | Cl$^-$ | I$^-$ | BF$_4^-$ | PF$_6^-$ | N(CN)$_2^-$ | C(CN)$_3^-$ | B(CN)$_4^-$ | FSI$^-$ | TFSI$^-$ | PFSI$^-$ | FEP$^-$ |
|---|---|---|---|---|---|---|---|---|---|---|---|---|
| Pyr$_4^+$ | 1.6 | 1.7 | 1.8 | 4.6 | 4.9 | 2.2 | 1.7 | 4.5 | 2.9 | 3.0 | 3.3 | 5.2 |
| AM$_4^+$ | 3.6 | 3.7 | 3.8 | 6.6 | 6.8 | 4.1 | 3.7 | 6.5 | 4.8 | 5.0 | 5.3 | 7.2 |
| IM$_4^+$ | 4.5 | 4.5 | 4.6 | 7.4 | 7.7 | 5.0 | 4.5 | 7.3 | 5.7 | 5.9 | 6.1 | 8.0 |
| Py$_4^+$ | 5.9 | 5.9 | 6.0 | 8.9 | 9.1 | 6.4 | 6.0 | 8.7 | 7.1 | 7.3 | 7.5 | 9.5 |

For all studied ionic liquids the difference between anionic and cationic $\varphi_S$ values ($\Delta\varphi_S$) fit into the predicted EW (Table 5). Considering the crudeness of approximations, that does not guarantee that the saturation potential is achievable within the measurable EW for all ions. Still, it is a good sign that saturation is experimentally verifiable.

An opposite conclusion applies to the possibility of fitting the monolayer potential into the EW (Table 4). The difference between anionic and cationic $\varphi_M$ values ($\Delta\varphi_M$) does not fit the predicted EW for most anion–cation combinations. An exceptional cation is Py$_n^+$ due to its flat geometry. As discussed above Py$_n^+$ is also the most polarisable among other studied cations. However, such polarisation reduces both $\varphi_M$ and the EW and, in sum, excludes monolayer formation prior to decomposition. Unlike Py$_n^+$, all exceptional anions (TFSI$^-$, PFSI$^-$, and FEP$^-$) are non-polarizable in simulations and electrochemically stable in experiments. Still, they are not as stable as predicted by Eq. (3) because of decomposition through bond breaking. Even if ILs consisting of TFSI$^-$, PFSI$^-$, or FEP$^-$ do not form the monolayer within real EW, these as well as BF$_4^-$, PF$_6^-$, B(CN)$_4^-$ anions are the best candidates for exploring the saturation regime at ionic liquid–electrode interfaces. However, ions of another shape should be considered to reach the monolayer formation.

*2.3. Ion shape and the monolayer potential*

The obtained data answers an essential question – whether monolayers of physisorbed ions are, in principle, achievable. The criteria for such a possibility is a realistically low $|\varphi_M|$ value below 2 V. For the ball ions, the potential can be translated into a radius of over 1.1 nm assuming that $\varepsilon_r = 2$ and $d = 0$ in Eq. (4). In other words, hard ball ions of ±1e charge and 1.1 nm radius form a monolayer at ∓2 V. All studied and most known ILs ions have much smaller dimensions. For comparison, the highest estimated $R$ value in this study equals 0.8 nm. An example from the literature is the polyoxometalate $PW_{12}O_{40}^{3-}$ anion with a radius of 0.55 nm. This anion was discussed by Borukhov *et al.* in Refs. [109,110] to illustrate the idea of crowding. A simple estimation of $\varphi_M$ for $PW_{12}O_{40}^{3-}$ gives an unreachable potential value of +13 V!

Figure 10 hints at an alternative way to lower the $|\varphi_M|$ by decreasing the compression factor. Indeed, a flat and wide frisbee-shaped cation with $r = 0.17$ nm and $R = 0.44$ nm should form a monolayer at −2 V. This is comparable to the size of 3*N*-coronene cations – one of the smallest representatives of nitrogen-substituted polycyclic aromatic hydrocarbon (PAH) cations [111]. Similar PAH could be substituted with boron [112] to provide frisbee anions. There is no record of PAH-based ILs being synthesised up to date. Yet, numerous experiments with triangulenium cations visualised ordered adlayers [113,114]. More realistic estimations at the density functional theory level predict that larger PAH ions are potential candidates for crowding at relatively small potentials [115].

## 2.4. Concluding remarks and future research directions

In this work, we performed classical MD simulations of the 40 most common ions (constituting ILs) on a graphene surface. Within this simple one-electrode model, we estimated the potentials of the saturated contact layer and monolayer: $\varphi_S$ and $\varphi_M$.

By comparing the estimated $\varphi_M$ and $\varphi_S$ values to predicted and literature data, it becomes evident that the monolayer and crowded structures of studied ions are unreachable within the experimentally measurable potential window. Still, reaching the crowded regime by designing ions of specific shape can reveal new phenomena of high applied potential, as predicted and discussed in Ref. [116]. This study shows that already-known phenomena (previously attributed to crowding) are due to the saturation of non-polarisable ions, *i.e.*, the formation of ordered layers at $\varphi_S$. Herewith, polarisable ions (like halides) might form the monolayer (exactly screening the surface charge) or even crowded EDL within the stability windows of corresponding ionic liquids due to partial charge transfer. From both fundamental and application perspectives, it is essential to study and understand the nature of adlayers formed by such polarisable as well as non-polarizable ions. The reported data sets the reference point for future studies, where theoretical $\varphi_S$ and $\varphi_M$ can be used as "milepost" potentials.

**2.5. Funding**

This work received financial support from FCT/MCTES (UIDP/50006/2020 DOI 10.54499/UIDP/50006/2020) through Portuguese national funds.

**2.6. Acknowledgements**

This work was supported by the Estonian Ministry of Education and Research (TK210). V.I. acknowledges the support of the Estonian Research Council (grant STP52). I.V.V. acknowledges funding from FCT/MCTES through the Portuguese national funds (LA/P/0008/2020 DOI 10.54499/LA/P/0008/2020, UIDP/50006/2020 DOI 10.54499/UIDP/50006/2020 and UIDB/50006/2020 DOI 10.54499/UIDB/50006/2020, REQUIMTE LAQV). Results were obtained using the High-Performance Computing Center of the University of Tartu.

**2.8. Supporting Information.**

The supporting information contains formulas and values for structure-determined potentials ($\varphi_S$ and $\varphi_M$) and capacitances ($C_S$ and $C_M$) in terms of a scaling exponent ($a$), the Helmholtz capacitance ($C_H$), relative and vacuum permittivities ($\varepsilon_r$ and $\varepsilon_0$), the distance between the EDL charge density planes ($l$, splitted into atomic radii $d$ and $r$) and the maximum packing density ($\theta_M$, recalculated into radius $R$). Codes for reproducing data and figures are accessible at https://github.com/vilab-tartu/NaRIBaS.